\documentclass[showkeys,showpacs,amssymb,nofootinbib]{revtex4}
\usepackage{amsmath}
\usepackage{amstext}
\usepackage{amsopn}
\usepackage{amsfonts}
\usepackage{amssymb}
\usepackage{accents}
\usepackage{empheq}
\usepackage{graphicx}
\usepackage{epsf}
\usepackage{graphics}
\usepackage{caption}
\usepackage{subcaption}
\usepackage[latin1]{inputenc}

\allowdisplaybreaks[3]
\def\F{F_{\mu\nu}}

\def\b{\begin{equation}}
\def\ee{\end{equation}}
\def\p{\partial}
\def\G{\Gamma}
\def\de{\delta}
\def\mn{\mu\nu}
\def\g{\mathcal{G}}
\begin{document}

\title{BRST renormalization of the first order Yang-Mills theory}

\author{J. Frenkel$^{a}$ and John C. Taylor$^{b*}$}
\affiliation{$^{a}$ Instituto de Física, Universidade de São Paulo, 05508-090,  São Paulo, SP, Brazil}
\affiliation{$^b$ Department of Applied Mathematics and Theoretical Physics, University of Cambridge, UK}

\footnote{Corresponding author *}
\footnote{\textit{E-mail addresses}: jfrenkel@if.usp.br (J. Frenkel), jct@damtp.cam.ac.uk (J.C. Taylor)}

\begin{abstract}
We examine the renormalization of the first order formulation of the Yang-Mills theory, by using the BRST identities. These preserve the gauge invariance of the theory and enable a recursive proof of renormalizability to higher orders in perturbation theory.  The renormalisation involves  non-linear mixings as well as re-scalings of the fields and sources, which lead to a renormalized action at all orders.
\end{abstract}
\keywords{gauge theories; first order; renormalization}

\pacs{11.15-q; 11.10.Gh}

\maketitle

\section{Introduction}
The first order formulation of gauge theories has a  simple form with  cubic 
interactions only \cite{okubo}.  The renormalization of the first order formalism has been previously studied  in the Yang-Mills theory, from various
points of view, mainly to one-loop order. This problem has been addressed in \cite{McKeon}, through the background filed quantization. The renormalization of a topological background Yang-Mills theory was examined in the Landau gauge in  \cite{Martellini}. A mathematical theorem concerning the renormalization of this formalism was derived in \cite{Costello}. The Hamiltonian formalism in the Coulomb gauge, which is first order in time derivatives, was studied using BRST in \cite{Andrasi}.

   What we add to the previous work is an account of how the BRST equation can be used to carry out renormalization to all orders of perturbation theory.  These identities reflect the gauge invariance of the theory and are suitable to prove, recursively, the renormalizability to higher orders in perturbation theory. In this formulation, the action depends upon the  bare
gluon field $A_{\mu}^B$    , the ghost fields $\eta^B,\overline{\eta}^B$             and on an additional field $F_{\mu\nu}^B$(colour indices are supressed). One can generate finite Greens functions by a rescaling of the gluon and ghosts fields as well as  of the coupling constant  ($A_{\mu},\eta,\overline{\eta}$ ,             and  $g$    are  renormalized quantities)                   
\begin{equation}
A_{\mu}^B=Z_A^{1/2}A_{\mu},  \:(\eta^B,\overline{\eta}^B)=Z_{\eta}^{1/2}(\eta,\overline{\eta}),\: g^B=Z'_g g.
\end{equation}
But the renormalization of the $\F^B$     field  requires, as well as scaling, a non-linear mixing with the gluon field
\begin{equation} 
F^B_{\mu\nu}=Z_F^{1/2}F_{\mu\nu}+Z_{FA}(\p_{\mu}A_{\nu}-\p_{\nu}A_{\mu})+gZ_{FAA}A_{\mu}\wedge A_{\nu}+.....
\end{equation}
which is admissible on dimensional and Lorentz symmetry grounds
(we use a color vector notation, with $A.B=A^a B^a$ and $(A\wedge B)^a=f^{abc}A^b B^c$, and leave the colour indices $a,b,c,..$ understood).   The ellipsis in (2) denotes a source term, which will be shown in equation (54).

 In the first order theory we introduce, for the purpose of setting up the BRST identities \cite{Taylor,Slavnov,Becchi, Kluberg}, 
 the standard Zinn-Justin sources $u_{\mu}^B$   and $v^B$,  and also a source  $w_{\mu\nu}^B$ for the gauge transformation of  the $\F^B$     field. In the renormalization procedure, these sources      also undergo a  re-scaling and nonlinear mixing which relate them to the renormalized sources   $u_\mu$, $v$   and $w_{\mu\nu}$. 
Although the above non-linear transformations are more involved than those which occur in the standard theory, we show that it is possible to renormalize this formulation to all orders, such that gauge invariance expressed through BRST identities is preserved order by order. 

In section 2 we define the basic  Lagrangian and give, in a general covariant gauge, the results for the one-loop ultra-violet divergent Green functions. In section 3 we introduce the BRST identities satisfied by the  renormalized action which generates the one-particle irreducible Green's functions. We also construct the appropriate counter-terms needed to renormalize the one-loop divergent functions. In section 4 we prove the renormalizability of this formulation to one-loop order, and also derive a basic connection between the bare and renormalized actions.  Using the BRST identities, we give in section 5 a recursive proof of renormalizability to all orders, which preserves to each order the gauge-invariance of the theory, and obtain the general renormalized Lagrangian (59).

To one-loop order, the divergences turn out to be simpler than the counter-terms which are allowed by BRST. In section 6, we explain why this occurs 
by exploiting the relationship between the first order formalism and the usual second order one, and show that this property persists to all orders. In consequence, the renormalized action is considerably simplified.

\section{The Lagrangian and one-loop divergences. }
The original Lagrangian density for the first order formulation in covariant gauges is
\b
\mathcal{L}'=\mathcal{L}_0-(1/2\xi)(\p^{\mu}A_{\mu})^2
\ee
where $\xi$ is a gauge-fixing parameter, and
\b
\mathcal{L}_0=\frac{1}{4}F_{\mu\nu}.F^{\mu\nu}-\frac{1}{2}F_{\mu \nu}.f^{\mu\nu} 
+(\p_{\mu}\overline{\eta}+u_{\mu}).D^{\mu}\eta -\frac{g}{2}v.(\eta\wedge\eta)+gw_{\mu\nu}.(F^{\mu\nu}\wedge \eta)
\ee

where $D_{\mu}$ is the covariant derivative and
\b
f_{\mu\nu}=\p_{\mu}A_{\nu}-\p_{\nu}A_{\mu}+gA_{\mu}\wedge A_{\nu}.
\ee
The coefficients of the auxiliary sources  in (4) are invariant under the BRST  gauge transformations 
\cite{Becchi}
\b
\delta F_{\mu\nu}=-g(F_{\mu\nu}\wedge \eta) \zeta,\,\, \delta A_{\mu}=-(D_\mu \eta)\zeta,\,\, \delta \eta=-\frac{g}{2}(\eta\wedge \eta )\zeta
\ee
which can be verified using the Jacobi identity ($\zeta$ is an infintesimal Grassmann quantity).

The Feynman rules obtained from (4) are given in the Appendix A and some examples of the evaluation of ultraviolet divergent graphs are given in  Appendix B. Here, we summarize only the resulting expressions for the one-loop divergent contributions from  the Feynman graphs. 
These contributions can be expressed in terms of the divergent constant (using dimensional
regularization in  $4-2\epsilon$   dimensions)
\b
d=\frac{g^2C_G}{16\pi^2 \epsilon}
\ee
The divergent parts of the $A$, $F$ self-energies and of the  $AF$ transition  diagrams are
\b
\Pi_{\mu\nu}^{ab}=\delta^{ab} d_{AA}(k_{\mu}k_{\nu}-\eta_{\mu\nu}k^2),   \  d_{AA}=\frac{1}{6}(-1+3\xi )d;
\ee
\b
\Pi^{ab}_{\alpha \beta,\alpha'\beta'}=\frac{1}{4}\delta^{ab}d_{FF}(\eta_{\alpha\alpha'}\eta_{\beta\beta'}-\eta_{\alpha\beta'}\eta_{\beta\alpha'}), \ d_{FF}=\frac{1}{2}(1+\xi)d;
\ee
\b
\Pi^{ab}_{\alpha\beta,\nu}=\frac{i}{2}\delta^{ab}d_{FA}(\eta_{\alpha\nu}k_{\beta}-\eta_{\beta\nu}k_{\alpha}),  \ d_{FA}=-\frac{1}{4}(3-\xi)d.
\ee
 The $AA$ and $FA$ amplitudes  in (8)  and (10) are transverse as a consequence of the BRST identities.
The divergent parts of the three-point $FAA$ and $AAA$  vertices have the forms (to one-loop order)
\b
V^{abc}_{\alpha\beta,\mu\nu}=\frac{g}{2}f^{abc}d_{FAA}(\eta_{\alpha\mu}\eta_{\beta\nu}-\eta_{\alpha\nu}\eta_{\beta\mu}), \ d_{FAA}=\frac{1}{2}\xi d;
\ee
\b
V^{abc}_{\mu\nu\rho}=-igf^{abc}d_{AAA}[\eta_{\mu\nu}(k_1-k_2)_{\rho}+\eta_{\nu\rho}(k_2-k_3)_{\mu}+\eta_{\rho\mu}(k_3-k_1)_{\nu}]
\ee
The divergent part of the one-loop four-gluon vertex has the structure
\b
V^{abcd}_{\lambda\mu\nu\rho}=g^2d_{AAAA}[(\eta_{\lambda\mu}\eta_{\nu\rho}-\eta_{\lambda\nu}\eta_{\mu\rho})f^{bce}f^{ade}] +\hbox{cyclic permutations of} \ (\mu,b),(\nu,c),(\rho,d).
\ee
 The coefficients $d_{AAA}$ and $d_{AAAA}$ have not been
calculated directly. Their value 
can be inferred indirectly from the BRST symmetry (see equation (35) below).
 
  The ultraviolet behaviour of the ghost sector  is the same as in the standard, second-order, theory. Thus, the divergent part of the ghost self energy, as well as the divergent parts of the graphs involving the sources $u$ and $v$ and external ghosts remain unchanged :
	\b
\frac{1}{4}d(-3+\xi)(\p^{\mu}\overline{\eta}+u^{\mu}).\p_{\mu}\eta+\frac{1}{4}\xi d[2g(\p^{\mu}\overline{\eta}+u^{\mu}).(A_{\mu}\wedge \eta)-gv.(\eta \wedge \eta)].
\ee	
The divergent contribution from one-loop graphs involving the source $w$ and external ghosts is
\b
\frac{d}{8}g\xi [2w^{\mu\nu}.(F_{\mu\nu}+f_{\mu\nu})\wedge \eta - (w^{\mu\nu}\wedge w_{\mu\nu}).(\eta \wedge \eta)]
\ee
which vanishes in the Landau gauge.
 
There is another possible independent structure
\b
C_1 w^{\mu\nu}.[(A_{\mu}\wedge \p_{\nu}\eta)-(A_{\nu}\wedge \p_{\mu}\eta)]
\ee
which by power-counting might be divergent; but explicit calculation of the one-loop Feynman graphs (see Appendix B) shows that its divergent part is zero. So, to one-loop order, no counter-term with the structure of (16) is necessary. In section 6, we explain why this cancellation occurs, and show it must happen to all orders.

In the next section, we show how to construct counter-terms to cancel the above divergences, consistently with gauge invariance.

\section{The one-loop counter-terms}
The original action is 
\b
\Gamma_0 = \int d^4 x \mathcal{L}_0(x).
\ee
Let $\Gamma$ be the complete effective action. The BRST  identities in the first order theory, which can be obtained in a similar manner to that used in the standard theory \cite{Taylor2}, are  
\b
\Gamma *\Gamma \equiv \int d^4 x \left[{\delta \Gamma \over \delta F_{\mu\nu}}.{\delta\Gamma \over w^{\mu\nu}}+{\delta \Gamma \over \delta A_{\mu}}.{\delta\Gamma \over \delta u^{\mu}}+{\delta \Gamma \over \delta\eta}.{\delta \Gamma \over \delta v}\right]=0
\ee
 These identities reflect the gauge invariance of the theory. For example, using (6) and (17), 
 (18)  to lowest order implies
\b
\int d^4x \left[{\delta\Gamma_0 \over \delta F_{\mu\nu}}.\delta F_{\mu\nu}+{\delta \Gamma_0 \over \delta A_{\mu}}.\delta A_{\mu}+{\delta \Gamma_0\over \delta\eta}.\delta \eta \right]=0
\ee
which states that $\Gamma_0$       is gauge invariant. 

Let   $\Gamma_1^{div}$ be the divergent part of the one-loop action, and $\Gamma_1^C=-\Gamma_1^{div}$                         be the counter-term action to one loop order. Then, from (18),
\b
\Delta\Gamma_1^C \equiv \Gamma_0 * \Gamma_1^C +\Gamma_1^C * \Gamma_0=0,
\ee
where
\b
\Delta = \int d^4x \left[ {\delta \Gamma_0 \over \delta F_{\mu\nu}}.{\delta \over \delta w^{\mu\nu}}+{\delta \Gamma_0 \over \delta w^{\mu\nu}}.{\delta \over \delta F_{\mu\nu}}+{\delta\Gamma_0 \over \delta A_{\mu}}.{\delta \over \delta u^{\mu}}+{\delta \Gamma_0 \over \delta u^{\mu}}{\delta \over \delta A_{\mu}}+{\delta \Gamma_0 \over \delta \eta}.{\delta \over \delta v}+ {\delta\Gamma_0 \over \delta v}.{\delta \over \delta \eta}\right].
\ee
It can be verified   that $\Delta$     is nilpotent:
\b
\Delta^2=0.
\ee

One class of solutions to equation  (20)     is of the form 
\b
\Gamma_1^{C1}=\Delta G,
\ee
where the  $G$ is a  polynomial in the fields and sources, which is a Lorentz scalar, which has ghost number 1, dimension of $(\hbox{mass})^{-1}$ and  which is invariant under rigid colour group transformations. The most general $G$ with these properties can be written as 
\pagebreak  
\b
G= \int d^4 x \bigg[\left(1-Z_A^{1/2} \right)A_{\mu}.u^{\mu}+ \left(1-Z_F^{1/2}\right)F_{\mu\nu}.w^{\mu\nu}-[Z_{FA}(\p_{\mu}A_{\nu}-\p_{\nu}A_{\mu})+gZ_{FAA}A_{\mu}\wedge A_{\nu}].w^{\mu\nu}$$
$$
-g\overline{Z}\eta.(w_{\mu\nu}\wedge w^{\mu\nu})+\left(Z_{\eta}Z_A^{1/2}-1\right)\eta .v \bigg]
\ee
Here the coefficients $Z_A, Z_F, Z_{\eta}$ will generate scaling and are of order $1+O(\hbar)$ to one-loop, and the other coefficients generate mixing and are $O(\hbar)$. The terms with coefficients $Z_{FA}$ and $Z_{FAA}$ are not gauge-invariant but they do meet the above criteria for inclusion in $G$. In general, they would generate the counter-term (16), but as we will show in  section 6, there is no divergent graph with this structure. The term (16) is absent if we choose
$Z_{FAA}=Z_{FA}$ at one-loop order. To all orders, the generalization of this condition turns out to be
\b
Z_{FAA}=Z'_gZ_{A}^{1/2}Z_{FA},
\ee
and henceforth we will impose this condition and eliminate $Z_{FAA}$ in favour of $Z_{FA}$. Using (25), equation  (23) generates the $Z_{FA}$ term in (29).
 (The reason for inserting the factor
$Z_A^{1/2}$ into the last term in (24) is to comply with conventional notation in the literature.)

A second type of solution of (20) consists of terms which are explicitly gauge invariant:
\b
\Gamma_1^{C2}= \int d^4x \left[\frac{1}{4}zF_{\mu\nu}.F^{\mu\nu}-\frac{1}{2}z'F_{\mu\nu}.f^{\mu\nu}-\frac{1}{4}z''f_{\mu\nu}f^{\mu\nu}\right],
\ee
the first two being proportional to terms in $\mathcal{L}_0$ in (4).

The third type of counter-term is obtained
by differentiating (18) with respect to the coupling constant g:  
\b
\Gamma_1^{C3}=Z'_g g\frac{d \Gamma_0}{dg},
\ee
so that $Z'_g$ represents a rescaling of the coupling constant.        

Combining the above three three contribution with  $\Gamma_0$     , we obtain to one-loop order
\b
\Gamma_0+\Gamma_1^{C1} +\Gamma_1^{C2} +\Gamma_1^{C3} = \int d^4 x [\mathcal{L}^i(x)+\mathcal{L}^{ii}(x)],
\ee
$\mathcal{L}^i$ and $\mathcal{L}^{ii}$ being the source-free and the source-containing parts respectively. These are as follows.
\b
\mathcal{L}^i=\frac{1}{4}(Z_F+z)F_{\mu\nu}.F^{\mu\nu}-\frac{1}{2}\left((Z_AZ_F)^{1/2}-Z_{FA}+z'\right)F_{\mu\nu}.\hat{f}^{\mu\nu}-\frac{1}{4}(2Z_{FA}+z'')f^{\mu\nu}.f_{\mu\nu}+Z_{\eta}\p_{\mu}\overline{\eta}.\hat{D}^{\mu}\eta,
\ee
\b
\mathcal{L}^{ii}=Z_{\eta}u_{\mu}.\hat{D}^{\mu}\eta+gZ_{\eta}Z'_g Z_A^{1/2}\left[w_{\mu\nu}.(F^{\mu\nu}\wedge \eta)-\frac{1}{2}v.(\eta\wedge \eta)\right]
+g\overline{Z}w_{\mu\nu}.(F^{\mu\nu}-f^{\mu\nu})\wedge \eta+\frac{1}{2}g^2\overline{Z}(w^{\mu\nu}\wedge w_{\mu\nu}).(\eta \wedge \eta)
\ee
Here we use the notation
\b
\hat{D}_{\mu}=\partial_{\mu}+gZ'_gZ_A^{1/2}A_{\mu}\wedge,\; \hat{f}_{\mu\nu}=\partial_{\mu}A_{\nu}-\partial_{\nu}A_{\mu}+gZ'_gZ_{A}^{1/2}A_{\mu}\wedge A_{\nu}.
\ee
As these  two equations have been written, they are correct only up to first order in $\hbar$, that is first order in $Z_{FA}, \overline{Z}, (Z_A-1), (Z_F-1), (Z_{\eta}-1), (Z'_g-1)$.

We will now show that the coefficients $z,z',z''$ are redundant, and we can choose them all to be zero. First, $z$ can be absorbed into $Z_F$,
then $z''$ can be absorbed into $Z_{FA}$; and finally $z'$ can be absorbed into $Z_A$, and also adjusting $Z'_g$ to keep $Z_A^{1/2}Z'_g$
unchanged. From now on we will put $z=z'=z^{\prime\prime}=0$.

We note here that the Lagrangian (29) may be  obtained from the source-free part of the tree Lagrangian (4), by substituting the renormalized fields by the bare ones given in eqs. (1)  and (2). In the next section it will be shown that, similarly, one can also obtain  the Lagrangian (30), by  making in  (4) appropriate re-scalings and mixings of the sources. 

\section{Renormalization to one-loop order and the  $\beta$     function. }

The renormalization is performed by requiring that the counter-terms cancel the divergences which arise in the evaluation of Feynman graphs. To this end we note that, since the ultra-violet ghost sector of the first order form is unchanged, it follows that $Z_{\eta}$     will be the same as  the 
ghost-field renormalization constant   conventionally called $\tilde{Z}_3$        in the standard second order formalism:
\b
Z_{\eta}=\tilde{Z}_3=1+\frac{1}{4}(3-\xi)d.
\ee
Moreover, from the behaviour of the one loop vertex   $\overline{\eta}A_{\mu}\eta$              (see appendix B) we then  obtain the relation
\b
Z'_gZ_A^{1/2}=Z_gZ_3^{1/2}
\ee
where $Z_g$     and  $Z_3^{1/2}$      are, respectively, the rescalings of the coupling constant and  of the gluon
field in the standard theory. Using these relations, we see that the last counter-term in (29) cancels
the divergent source-free ghost terms in (14).

In order for the counter-terms in (29) to cancel other
one-loop divergences coming from the source-free graphs, we require the conditions
\b
Z_F-1=-d_{FF},
\ee

\b
-2Z_{FA}=d_{AA}=d_{AAA}=d_{AAAA},
\ee
\b
Z'_g-1=\frac{3}{2}d_{AA}-\frac{1}{2}d_{FF}+2d_{FA}-d_{AAA}-d_{FAA},
\ee
\b
Z_A-1=-d_{AA}+d_{FF}-2d_{FA},
\ee
where the divergent terms on the right hand sides of these equations were given in section 2, except $d_{AAA}$ and $d_{AAAA}$,
whose values are only inferred.

In order to make a further comparison with the second order formulation, we note that Green functions are the same
in both formulations, which follows from integrating out the $F_{\mu\nu}$ field. As shown in Appendix C, this implies that 
\b
-d_{AA}+d_{FF}-2d_{FA}=Z_3 -1.
\ee
Then, comparing with (37), and using (33), we obtain 
\b
Z_A=Z_3=1+\frac{d}{6}(13-3\xi),
\ee
\b
Z'_g=Z_g=1-\frac{11}{6}d,
\ee
leading to the expected result for the $\beta$-function, which is responsible for asymptotic freedom.

We now consider the renormalization of the graphs involving the sources.  The terms involving the sources  $u$  and $v$   in  (14)  are  canceled by corresponding 
terms  from (30), because of the relation
\b
Z_{\eta}+Z'_g +\frac{1}{2}Z_A -\frac{5}{2}=-\frac{1}{2}\xi d.
\ee
The remaining counter-term is $\overline{Z}$, and this is determined by comparing (15) with (30) to be
\b
\overline{Z}=\frac{1}{4}d\xi .
\ee
Note that both (41) and (42) vanish in the Landau gauge.

Finally, to complete the analysis to one loop order, we show that the counter-terms in (29) and (30) can be obtained by re-scaling and mixing both the fields (as in (1) and (2)) and the sources. The necessary transformations may be written
\b
F_{\mu\nu}\rightarrow F^B_{\mu\nu}=Z_F^{1/2}F_{\mu\nu}+Z_{FA}f_{\mu\nu}-2g\overline{Z}w_{\mu\nu} \wedge \eta,
\ee
\b
w_{\mu\nu}\rightarrow w^B_{\mu\nu} =(Z_A Z_{\eta}/Z_F)^{1/2} w_{\mu\nu},
\ee
\b
v\rightarrow v^B =Z_A^{1/2}v-g\overline{Z}w_{\mu\nu} \wedge w^{\mu\nu},
\ee
\b
u_{\mu} \rightarrow u^B_{\mu}= Z_{\eta}^{1/2}u_{\mu}-2Z_{FA}D^{\nu}w_{\mu\nu}.
\ee
(Although we have written these transformations with products of $Z$ factors, only the contributions up to order $\hbar$ are relevant at this stage.) With equation (1) and the above four transformations, we have that
\b
\Gamma^R_{(1)} \equiv (\Gamma_0+\Gamma_1^C)(g;A,\eta, \overline{\eta},F;u,v,w)=\Gamma_0(g^B;A^B,\eta^B,\overline{\eta}^B,F^B ;u^B,v^B,w^B).
\ee
Here the suffices $0$ and $1$ denote zeroth and first order in $\hbar$, $(1)$ denotes up to and including first order, and superfix $B$ denotes 'bare'  quantities.

We have thus renormalized to one loop the first order formulation of the theory, by rescaling/mixing the bare coupling , fields and sources in a way consistent with the BRST  identities.

\section{Renormalization to all orders}
The previous procedure for renormalizing the first order Yang-Mills theory may be extended, 
recursively, to higher orders in perturbation theory. One must show that we can rescale/mix the  coupling constant, the fields and the sources so that the effective action contains finite Green 
functions and BRST invariance is maintained at each order. We have seen that this works to
one loop order. The proof that this holds to higher orders may be made by induction, using a
reasoning somewhat similar to that employed in the standard theory \cite{Taylor2, Itzykson}. However, due to
the non-linear mixing of the fields and sources, some arguments are more involved.  
In order for this recursive procedure to work, it is necessary to show that the renormalised 
action, up to and including order $\hbar^N$,
\b
 \Gamma^R_N =\Gamma_0+\Gamma^C_N
\ee
satisfies
\b
\Gamma^R_N * \Gamma^R_N=0.
\ee

In the recursive procedure to go to the next order, one will add terms of the form
\b
\Delta_NG_{(N+1)}
\ee
where $\Delta_N$ is defined like in equation (21) but with $\Gamma^R_N$ replacing $\Gamma_0$, and $G_{(N+1)}$ is the $(N+1)$th order contribution to (24). However,
this does not insure that
\b
\Gamma^R_N+\Delta_N G_{(N+1)}
\ee
satisfies a BRST condition analagous to (49). We need to prove that there is a modification of (51) which does satisfy BRST.
This problem appears to be harder than in conventional second-order QCD, because terms of third degree appear in $G$ (see equation (24)),
leading to non-linear mixing, as in (29) and (30). Nevertheless, we give  a theorem which provides what is needed.

The theorem states that
\b
\Gamma_0 +\Gamma^C \equiv \Gamma^R(g;A,F,\eta,\overline{\eta};u,v,w)=\Gamma_0(g^B;A^B,F^B,\eta^B,\overline{\eta}^B;u^B,v^B,w^B)
\ee
satisfies BRST provided that the bare quantities are related to the ordinary ones by the re-scaling and (non-linear) mixing
equations:

\b
g^B=Z'_g g, \; A^B_{\mu}=Z_A^{1/2}A_{\mu}, \; (\eta^B,\overline{\eta}^B)=Z_{\eta}^{1/2}(\eta, \overline{\eta}),
\ee
\b
F^B_{\mu\nu}=Z_F^{1/2}F_{\mu\nu}+Z_{FA}(\p_{\mu}A_{\nu}-\p_{\nu}A_{\mu})+gZ_{FAA}A_{\mu}\wedge A_{\nu}-2gZ_F^{-1/2}\overline{Z}w_{\mu\nu}\wedge \eta,
\ee
\b
u_{\mu}^B=Z_{\eta}^{1/2}u_{\mu}-2(Z_{\eta}/Z_F)^{1/2}(Z_{FA}\p^{\nu}w_{\mu\nu}+gZ_{FAA}A^{\nu}\wedge w_{\mu\nu}),
\ee
\b
w^B_{\mu\nu}=(Z_AZ_{\eta}/Z_F)^{1/2}w_{\mu\nu}, \; v^B=Z_A^{1/2}\,v-g(Z_A^{1/2}\overline{Z}/Z_F)w_{\mu\nu}\wedge w^{\mu\nu}.
\ee
Here the equations have been written to infinite order, but in the iteration process they could be truncated at any required order.
The proof of this theorem is given in Appendix D.

In (54) and (55), we have not imposed the condition (25). When we do so, these equations simplify to

\b
F^B_{\mu\nu}=Z_F^{1/2}F_{\mu\nu}+Z_{FA}\hat{f}_{\mn}-2gZ_F^{-1/2}\overline{Z}w_{\mu\nu}\wedge \eta,
\ee
\b
u_{\mu}^B=Z_{\eta}^{1/2}u_{\mu}-2(Z_{\eta}/Z_F)^{1/2}Z_{FA}\hat{D}^{\nu}w_{\mn},
\ee
using the notation of equation (31).

When (53),(54), (57), (58) are inserted into (52), we obtain the all orders renormalized Lagrangian
\b
\mathcal{L}^R=\frac{1}{4}Z_FF^{\mu\nu}.F_{\mu\nu}+\frac{1}{2}Z_F^{1/2}\left(Z_{FA}-Z_A^{1/2}\right)F^{\mu\nu}.\hat{f}_{\mu\nu}+\frac{1}{4}Z_{FA}\left(Z_{FA}-2Z_A^{1/2}\right)\hat{f}^{\mu\nu}.\hat{f}_{\mu\nu}$$
$$+Z_{\eta}(\p^{\mu}\overline{\eta} +u^{\mu})\hat{D}_{\mu} \eta+gZ'_gZ_A^{1/2}Z_{\eta}\left[w^{\mu\nu}.(F_{\mu\nu}\wedge \eta)-\frac{1}{2}v.(\eta\wedge\eta)\right]$$
$$+g\overline{Z}w^{\mu\nu}.\left[\left\{F_{\mu\nu}+\left(Z_{FA}-Z_A^{1/2}\right)Z_F^{-1/2}\hat{f}_{\mu\nu}\right\}\wedge \eta \right]+\frac{1}{2}g^2\overline{Z}Z_F^{-1}\left(Z_{\eta}Z'_gZ_A^{1/2}-\overline{Z}\right)(w^{\mu\nu} \wedge w_{\mu\nu}).(\eta \wedge \eta).
\ee
Renormalization would proceed by expanding the coefficients in (59) in powers of $\hbar$, and, to $n$-loop order, adjusting the
$\hbar^n$ terms to cancel the divergences. 

 Note that the gauge-transform of $A$ implied by the $u$ term in (59) has the same form as (6),  for the original Lagrangian (4); but the transformation of $F$ implied by the $w$ term is inhomogeneous and more complicated.

We argue that counter-terms like in equation (26) can be omitted to all orders. Take for example the term $z''\hat{f}_{\mu\nu}.\hat{f}^{\mu\nu}$ which is gauge-invariant under the gauge transformation of $A$ implied by the fourth term in (59). This can be absorbed into the third term in (59) by adjusting $Z_A$ and $Z_{FA}$, and at the same time adjusting $Z_F$ and $Z'_g$ so as to keep all the other coefficients in (59) fixed.

The last line of (59), involving $\overline{Z}$, has no physical consequence (and, according to (42) is zero in the Landau gauge).
If we ignore this line, the remainder of (59) is invariant under gauge-transformations of the same form as those in (6) for the original Lagrangian (4). This is evident from the second line of (59). 

\section{A restriction on the divergences}
In this section, we prove, to all orders, that the term in (16) is not divergent. This implies equation (25)
and thus reduces the number of counter-terms needed. Our method is to take advantage of the fact that the Lagrngian (4) is quadratic in
$F$, and so this can be integrated out, leaving the ordinary, second order Yang-Mills Lagrangian. The source $w$ remains, but is now an inert
external source, not changing under gauge transformations. It turns out that the BRST identity for the resulting theory is more restrictive than
(18), and forbids a divergence of the form (16).

Using functional integrals, the generating functional $G(J,\phi, \overline{\phi})$ for connected Green functions is given by
\b
\exp G \propto \int dFdAd\eta d\overline{\eta} \exp[\Gamma_0+J_{\mu}.A^{\mu}+\overline{\phi}.\eta+\overline{\eta}.\phi],
\ee.
where $\Gamma_0 =\int d^4x \mathcal{L}_0$ in (4). It would have been possible to include a source for $F_{\mn}$ in (60), but for present purposes
we omit it. The functional integration over $F_{\mn}$ may be carried out, giving
\b
\exp G\propto \int dAd\eta d\overline{\eta} \exp[\Gamma'_0+J_{\mu}.A^{\mu}+\overline{\phi}.\eta+\overline{\eta}.\phi],
\ee
where $\Gamma'_0$ comes from $\mathcal{L}'_0$ given by
\b
\mathcal{L}'_0=-\frac{1}{4}f_{\mn}.f^{\mn}+(\p_{\mu}\overline{\eta}+u_{\mu}).D^{\mu}\eta -\frac{g}{2}v.(\eta \wedge \eta)+gw_{\mn}.(f^{\mn}\wedge \eta)-g^2(w_{\mn}\wedge \eta).(w^{\mn}\wedge\eta).
\ee
This Lagrangian is invariant under the gauge transformations
\b
\delta A_{\mu}=-(D_{\mu}\eta)\zeta,\,\,\,\delta \eta=-\frac{1}{2}g(\eta \wedge \eta)\zeta.
\ee
which may be compared to (6). The external source $w^a_{\mn}$ does \emph{not} transform. (To avoid any confusion, we note that (61) is invariant under rigid colour transformations, in which $A, f, c, w$ all transform by space-independent matrices of the adjoint representation of the colour group.) So $\Gamma'_0$ satisfies the BRST condition
\b
\Gamma'_0 *' \Gamma'_0 \equiv\int d^4x\left[\frac{\delta \Gamma_0'}{\delta A_{\mu}}.\frac{\delta\Gamma'_0}{\delta u^{\mu}}+\frac{\delta \Gamma_0'}{\delta \eta}.\frac{\delta\Gamma'_0}{\delta v}\right]=0,
\ee
where the symbol $*'$ signifies that only the two terms appear in (64), rather than the three in (18).

Let $\Gamma'(A,\eta,\overline{\eta};u,v)$ be the effective action derived from $\Gamma'_0$. Then, by the standard arguments, this satisfies
\b
\Gamma' *' \Gamma'=0.
\ee
To one-loop order, this gives (analagously to (20))
\b
\Gamma'_0 *' \Gamma_1^{'C}+\Gamma_1^{'C}  *' \Gamma'_0=0,
\ee
implying that the counter-terms $\Gamma_1^{'C}$ are gauge-invariant under (63). The term (16) is not so invariant, and so is forbidden, and the corresponding divergence must be zero. This explains the result found in Appendix B from the graphs in Fig.3.

To extend this argument to all loops, we must first allow for the addition of counter-terms to $\Gamma_0$ and consequently to $\Gamma'_0$. According to (52), this is taken account of by replacing all the fields and sources and $g$ by bare quantities, $A^B$,..etc. Then, instead of integrating over $F$, we integrate over $F^B$. The result is 
\b
\Gamma'(g^B;A^B,\eta^B,\overline{\eta}^B;u^B,v^B;w^B),
\ee
which, according to the theorem in Appendix D, satisfies the BRST condition (65).

We can now prove that the term (16) is forbidden to all orders. Suppose there was a non-zero contribution to it at $n$-loop order
but not to lower order. Then, taking the BRST condition (66) to order $\hbar^n$, and picking out the counter-term (16), we get 
\b
\Gamma'_0 *' \Gamma_n^{'C}+\Gamma_n^{'C}  *' \Gamma'_0=0,
\ee
and, just as with (66), this would imply that (16) is invariant under (63), a contradiction.
\section{Conclusion}
We have studied, in a general covariant gauge, the renormalization of the first order formalism of Yang-Mills theory. We have shown that, to all orders, the renormalization can be controlled by the BRST
identities, so as to preserve gauge invariance. The bare and renormalized fields and Zinn-Justin sources
are related by non-linear mixing as well as scaling. A term which is allowed within the first order  formalism is shown to be zero, when the  relation to the second order formalism is exploited.
\appendix
\renewcommand{\theequation}{A.\arabic{equation}}
\setcounter{equation}{0} 
\section*{Appendix A}
The Feynman rules for the first order Yang-Mills theory \cite{Brandt} are shown in Fig.1. Here, the tensors   $I_{\lambda\sigma,\rho\kappa}$ and $L_{\lambda\sigma,\rho\kappa}$                                are  given in momentum space by 
\b
I_{\lambda\sigma,\rho\kappa}=\frac{1}{2}(\eta_{\lambda\rho}\eta_{\sigma\kappa}-\eta_{\lambda\kappa}\eta_{\sigma\rho}),\label{A1}
\ee
\b
L_{\lambda\sigma,\rho\kappa}(p)=\frac{1}{2}(p_{\lambda}p_{\rho}\eta_{\sigma\kappa}+p_{\sigma}p_{\kappa}\eta_{\lambda\rho}-p_{\lambda}p_{\kappa}\eta_{\sigma\rho}-p_{\sigma}p_{\rho}\eta_{\lambda\kappa}).
\ee
 Note that the $F$-propagator is transverse and the  $AF$-propagator is transverse on its $\mu$ index.
\begin{figure}[h]
\centering
\includegraphics[width=10cm]{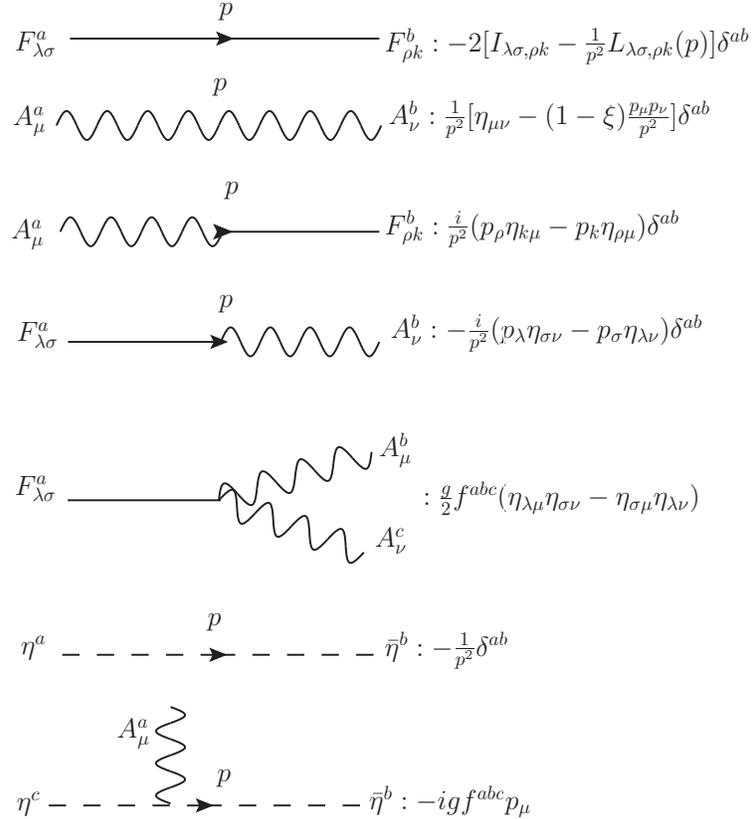}
\caption{Feynman rules for the propagators and vertices in covariant gauges. Ghosts are denoted by dashed lines. 
\newline
($\xi$ is a parameter defining the gauge).}
\end{figure}
\renewcommand{\theequation}{B.\arabic{equation}}
\setcounter{equation}{0}
\section*{Appendix B}
Here we give two examples of the evaluation of ultraviolet divergent parts of Green functions 
We first evaluate the graphs contributing to the ghost-gluon vertices shown in Fig.2.
The contribution from the first Feynman graph is
\b
V_{1\mu}^{abc}=-ig^3f^{ac'b'}f^{a'bc'}f^{a'b'c}\int \frac{d^n p}{(2\pi)^n i}\frac{(p+q')_{\mu}}{(p+q')^2}\frac{(p+q)_{\lambda}}{(p+q)^2}q'_{\nu}\frac{1}{p^2}\left(\eta^{\nu\lambda}-(1-\xi)\frac{p^{\nu}p^{\lambda}}{p^2}\right).
\ee
Using $f^{ac'b'}f^{a'bc'}f^{a'b'c}=(C_G/2)f^{abc}$, and keeping only terms which contribute to the ultra-violet divergence, we get
\b
V_{1\mu}^{abc}=-ig^3(C_G/2)\xi f^{abc}\int \frac{d^n p}{(2\pi)^n i}q'^{\nu}\frac{p_{\mu}p_{\nu}}{(p+q)^2 p^2(p+q')^2 }
\ee
Taking $n=4-2\epsilon$, the divergent part is
\b
-ig^3f^{abc}q'_{\mu}\frac{C_G}{16\pi^2}\xi \frac{1}{8\epsilon}.
\ee

\begin{figure}
\centering
\begin{subfigure}[b]{0.3\textwidth}
        \includegraphics[width=\textwidth]{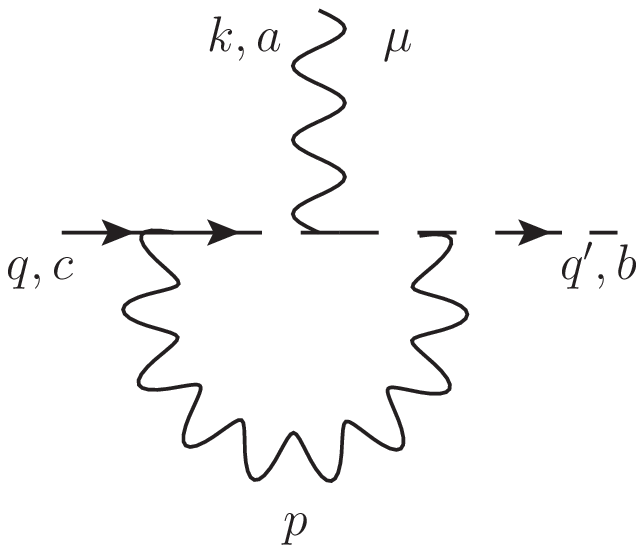}
        \caption{\label{fig:fig1}}
\end{subfigure}
\begin{subfigure}[b]{0.3\textwidth}
        \includegraphics[width=\textwidth]{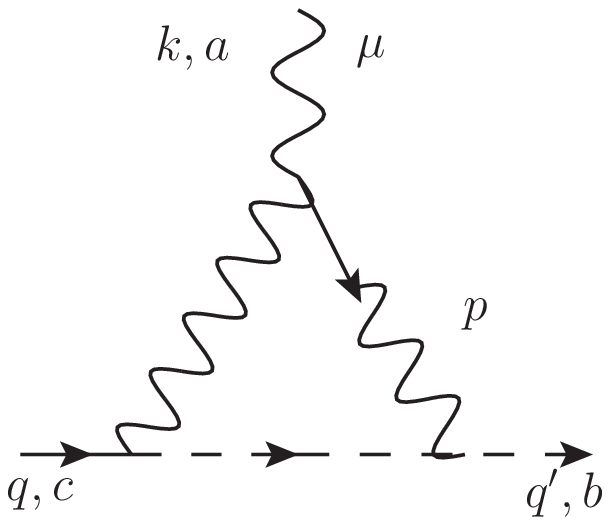}
        \caption{\label{fig:fig2}}
       \end{subfigure}
\begin{subfigure}[b]{0.3\textwidth}
        \includegraphics[width=\textwidth]{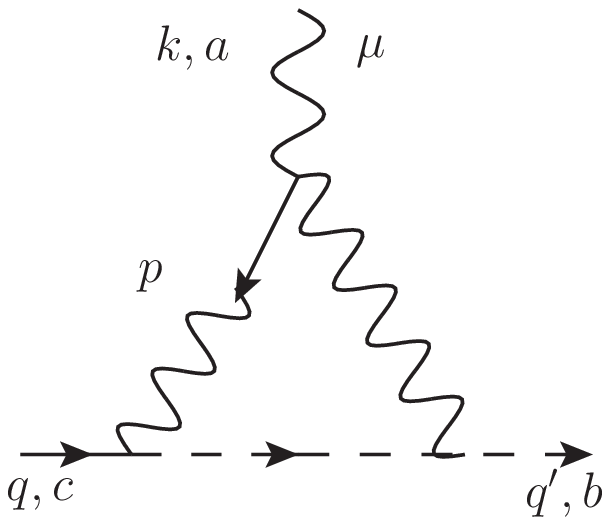}
        \caption{\label{fig:fig3}}
\end{subfigure}
\caption{One-loop contributions to the ghost-gluon vertex.}
\end{figure}
Similarly, graph(b) contributes
\b
-ig^3f^{abc}q'_{\mu}\frac{C_G}{16\pi^2}\xi \frac{3}{8\epsilon}.
\ee
Graph (c) is UV finite, so the total is
\b
-ig^3f^{abc}q'_{\mu}\frac{C_G}{16\pi^2}\xi \frac{1}{2\epsilon},
\ee
the same as in standard, second order, Yang-Mills theory.

The second example is the graphs in Fig.3, which contribute to the ghost-gluon-$w$ vertex.
\begin{figure}
\centering
\begin{subfigure}[b]{0.3\textwidth}
        \includegraphics[width=\textwidth]{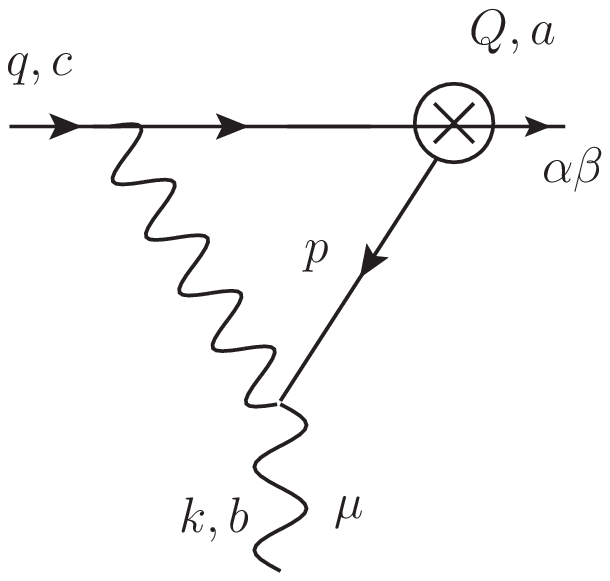}
        \caption{\label{fig:fig1}}
\end{subfigure}
\begin{subfigure}[b]{0.3\textwidth}
        \includegraphics[width=\textwidth]{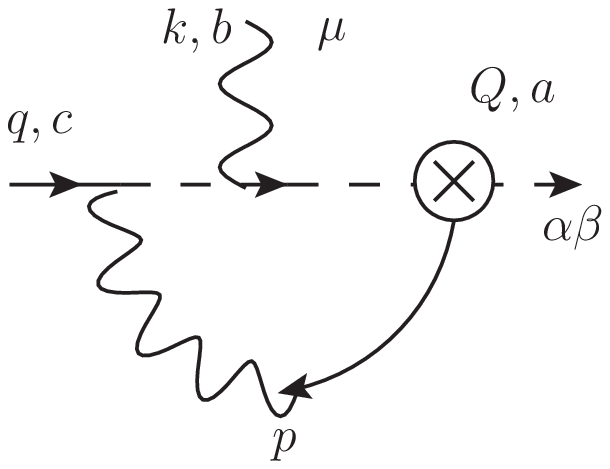}
        \caption{\label{fig:fig2}}
       \end{subfigure}
\begin{subfigure}[b]{0.3\textwidth}
        \includegraphics[width=\textwidth]{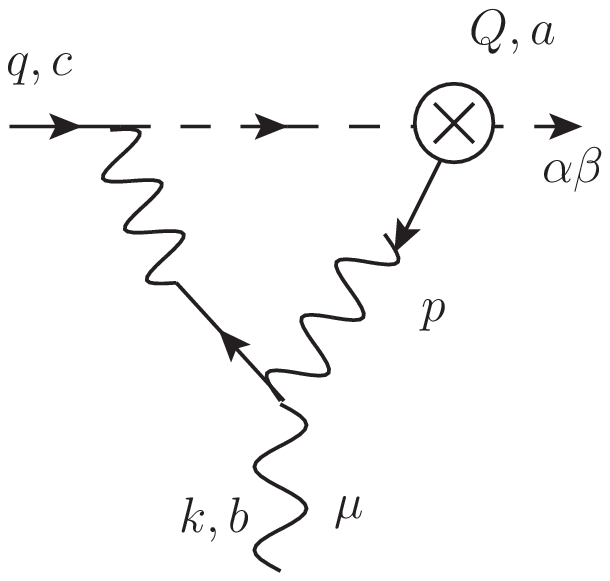}
        \caption{\label{fig:fig3}}
\end{subfigure}
\caption{One-loop contributions to the ghost-gluon-$w$ vertex. The source $w$ is denoted by a cross.}
\end{figure}
We seek to calculate the divergent coefficient $C_1$ in (16).
The $wF\eta$ vertex (denoted by a cross) is read off from the Lagrangian $\mathcal{L}_0$ in equation (4).
To find the divergent parts, we may set $k=0$. The contributions from the three graphs are proportional to $\frac{1}{4},-\frac{1}{8},-\frac{1}{8}$, giving a zero total. This result is independent of the gauge parameter $\xi$. The reason is because the $\xi$
term is proportional to $p^{\sigma}$, which gives zero because the $F$ propagator is transverse.
\section*{Appendix C}
\renewcommand{\theequation}{C.\arabic{equation}}
\setcounter{equation}{0}
The purpose of this section is to derive the relation (38).  The relevant one-loop diagrams are shown in Fig.4.
\begin{figure}[h]
\centering
\includegraphics[scale=0.68]{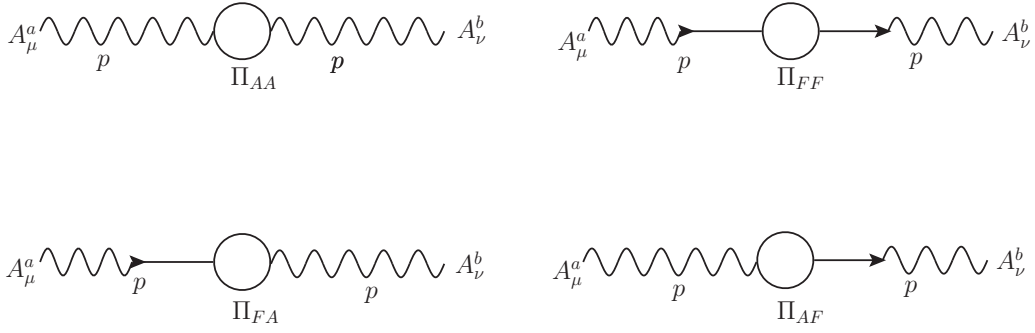}
\caption{One-loop contributions to the gluon propagator $D_{\mu\nu}^{ab}(p)$.}
\end{figure}
Using the divergent parts listed in section 2 and the Feynman rules in Appendix A, the divergent contributions from the four graphs give
\b
\delta_{ab}\frac{1}{(p^2)^2}(d_{AA}-d_{FF}+2d_{FA})(p_{\mu}p_{\nu}-\eta_{\mu\nu}p^2).
\ee
 In the standard, second order, Yang-Mills theory,
there is a single graph with the same $p$-dependence, and a divergent coefficient conventionally  called $1-Z_3$. Comparison yields
equation (38).

\section*{Appendix D}
\renewcommand{\theequation}{D.\arabic{equation}}
\setcounter{equation}{0}

The purpose of this appendix is to verify that the transformations (53),...(56) preserve the BRST condition. Calculating the three terms in BRST, we find:
\b
\frac{\delta\Gamma_R}{\de A_{\mu}}.\frac{\de\G_R}{\de u^{\mu}}=(Z_AZ_{\eta})^{1/2}\frac{\delta\Gamma_R}{\de A^B_{\mu}}.\frac{\delta \Gamma_R}{\delta u^{B\mu}}
+2Z_{\eta}Z_F^{-1/2}Z_{FAA}w_{\mn}.\left(\frac{\de\Gamma_R}{u_{\mu}^B}\wedge \frac{\de\G_R}{\de u_{\nu}^B}\right)$$
$$+2Z_{\eta}^{1/2}
\frac{\de\G_R}{\de u^{B\mu}}.\left[Z_{FA}\p{\nu}\frac{\de\G_R}{\de F_{\mn}^B}-gZ_{FAA}\frac{\de \Gamma_R}{\de F_{\mn}^B}\wedge A_{\nu}\right],
\ee
\b
\frac{\delta\G_R}{\delta  F_{\mn}}.\frac{\delta\G_R}{\de w^{\mn}}=(Z_AZ_{\eta})^{1/2}\frac{\de \G_R}{\de F_{\mn}^B}.\frac{\delta \G_R}{\delta w^{B\mn}}+2gZ_F^{-1/2}\overline{Z}\eta.\left(\frac{\de\G_R}{\delta F^B_{\mn}}\wedge \frac{\de \G_R}{\delta F^{B\mn}}\right)
$$
$$+\frac{\delta\G_R}{\de F^B_{\mn}}.\left[2Z_{\eta}^{1/2}\left(Z_{FA} \p_{\nu}\frac{\de\G_R}{\de u^{B\mu}}-gZ_{FAA}\frac{\de \G_R}{\de u^{B\mu}} \wedge A_{\nu}\right)-2g(Z_A/Z_F)^{1/2}\overline{Z}\frac{\de \G_R}{\de v^B}\wedge w^B_{\mn}\right],
\ee
\b
\frac{\de\G_R}{\de \eta}.\frac{\de\G_R}{\de v}=(Z_AZ_{\eta})^{1/2}\frac{\de\G_R}{\de \eta^B}.\frac{\de\G_R}{\de v^B}+2g(Z_A/Z_F)^{1/2}\overline{Z}\frac{\de \G_R}{\de F^B_{\mn}}.\left(\frac{\de\G_R}{\de v^B}\wedge w^B_{\mn}\right).
\ee
In the sum of these three equations, the second terms in (D.1) and (D.2) vanish by antisymmetry. The last two terms in (D.1) cancel with the two terms next to the last in (D.2), and the last two terms in (D.2) and (D.3) cancel similarly. Thus the total gives
\b
\frac{\de \G_R}{\de A_{\mu}}.\frac{\de \G_R}{\de u^{\mu}}+\frac{\de \G_R}{\de F_{\mn}}.\frac{\de \G_R}{\de w^{\mn}}+\frac{\de \G_R}{\de \eta}.\frac{\de \G_R}{\de v}$$
$$ =(Z_AZ_{\eta})^{1/2}\left[\frac{\de \G_R}{\de A^B_{\mu}}.\frac{\de \G_R}{\de u^{B\mu}}+\frac{\de \G_R}{\de F^B_{\mn}}.\frac{\de \G_R}{\de w^{B\mn}}+\frac{\de \G_R}{\de \eta^B}.\frac{\de \G_R}{\de v^B}\right].
\ee
Then, as the right hand side of (52) satisfies the BRST condition (18), it follows that the left hand side does too.

The factor $(Z_AZ_{\eta})^{1/2}$ above is an artefact of our particular definition of the bare fields, and could be removed  (using ghost number conservation) by using new bare ghost and sources 
\b
u^{B'}_{\mu}=(Z_AZ_{\eta})^{-1/2}u^B_{\mu},\;\,\, w^{B'}_{\mn}=(Z_AZ_{\eta})^{-1/2}w^B_{\mn}, \,\; \eta^{B'}=(Z_AZ_{\eta})^{1/2}\eta^B,\;\,
v^{B'}=(Z_AZ_{\eta})^{-1}v^B,
\ee
which  would  not change $\G_R$. 

In conclusion, we observe that the all-order transformations (53) to (56), can be produced from the generating function $G$ in (24) by a modification of the 
lowest-order equation (23). The method is based on an analogy with classical mechanics, where canonical transformations are produced from generating functions.

Define
\b
\overline{G}(A,F,\eta;u,w,v)=-A_{\mu}.u^{\mu}-F_{\mn}.w^{\mn}+\eta.v+G,
\ee
where $G$ was defined in (24). Then we shall use
\b
\g=\overline{G}(A,F,\eta;u^{B'},w^{B'},v^{B'}),
\ee
with the bare sources defined in (D.5). Then the following equations give implicitly the transformations (53) to (56):
\b
\frac{\de\g}{\de A_{\mu}}=-u^{\mu},\,\, \frac{\de\g}{\de F_{\mn}}=-w^{\mn},\,\,\frac{\de\g}{\de \eta}=-v;\,\,\frac{\de\g}{\de u_{\mu}^{B'}}=A^{B'\mu},\,\,\frac{\de\g}{\de w^{B'}_{\mn}}=F^{B'\mn},\,\,\frac{\de\g}{\de v^{B'}}=\eta^{B'}.
\ee

\section*{Acknowledgment}
J.F. would like to thank F.T.Brandt and A.L.M. Britto for helpful conversations, and CNPq (Brazil) for a grant.

\end{document}